\documentclass[12pt]{article}
\usepackage{amssymb}
\begin{document}
\begin{center}
The nimbus of away-side jets \\

I.M. Dremin\\ 

{\it Lebedev Physical Institute, Moscow, Russia}

\end{center}

\begin{abstract}
The conical structure around the away-side jets is discussed.
The equations of in-medium gluodynamics are proposed. Their 
classical lowest order solution is explicitly shown for a color charge
moving with constant speed. For nuclear permittivity larger than 1 it
describes the shock wave induced by emission of Cherenkov gluons. 
The values of real and imaginary parts of nuclear permittivity are estimated 
from fits of RHIC data. Specific effects at LHC energies are described.
\end{abstract}

The conical structure around away-side jets has been observed in high-energy
central nucleus-nucleus collisions at RHIC \cite{ajit, ul, pru}. It can be 
explained
as the emission of Cherenkov gluons by a parton passing through a quark-gluon 
medium. The properties and evolution of the medium are widely debated. At the 
simplest level it is assumed to consist of a set of current quarks and gluons. 
The collective excitation modes of the medium may, however, play a crucial role.
Phenomenologically their impact would be described by the nuclear permittivity
of the matter corresponding to its response to passing partons. Namely this
approach is most successful for electrodynamical processes in matter.
Therefore, it is reasonable to modify the QCD equations by taking into account
collective properties of the quark-gluon medium \cite{depj}. Strangely enough,
this was not done earlier. For the sake of simplicity we consider here the 
gluodynamics only.

The classical lowest order solution of these equations coincides with
Abelian electrodynamical results up to a trivial color factor. One of the most
spectacular of them is Cherenkov radiation and its properties. Now, Cherenkov
gluons take the place of Cherenkov photons \cite{d1, ko}. Their emission in 
high-energy hadronic collisions is described by the same formulae but 
with the nuclear permittivity in place of the usual one. Actually, one 
considers them as quasiparticles, i.e. quanta of the medium excitations 
leading to shock waves with properties determined by the permittivity. 

Another problem of this approach is related to the notion of the
rest system of the medium. It results in some specific features of this
effect at LHC energies.

To begin, let us recall the classical in-vacuum Yang-Mills equations
\begin{equation}
\label{f.1}
D_{\mu}F^{\mu \nu }=J^{\nu }, \;\;\;\;\;
F^{\mu \nu }=\partial ^{\mu }A^{\nu }-\partial ^{\nu }A^{\mu }-
ig[A^{\mu },A^{\nu }],
\end{equation}
where $A^{\mu}=A_a^{\mu}T_a; \; A_a (A_a^0\equiv \Phi_a, {\bf A}_a)$ are the 
gauge field (scalar and vector) potentials, the color matrices $T_a$ satisfy
the relation $[T_a, T_b]=if_{abc}T_c$, $\; D_{\mu }=\partial _{\mu }-ig[A_{\mu }, \cdot], \;\; 
J^{\nu }(\rho, {\bf j})$ a classical source current, the 
metric $g^{\mu \nu }$=diag(+,--,--,--).

The chromoelectric and chromomagnetic fields are
$E^{\mu}=F^{\mu 0 }, \;
B^{\mu}=-\frac {1}{2}\epsilon ^{\mu ij}F^{ij}$
or, as functions of the gauge potentials in vector notation,
\begin{equation}
\label{4}
{\bf E}_a=-{\rm grad }\Phi  _a-\frac {\partial {\bf A}_a}{\partial t}+
gf_{abc}{\bf A}_b \Phi _c, \;\;\;\;
{\bf B}_a={\rm curl }{\bf A}_a-\frac {1}{2}gf_{abc}[{\bf A}_b{\bf A}_c].
\end{equation}

Herefrom, one easily rewrites the in-vacuum equations of motion (\ref{f.1}) in 
vector form. We do not show them explicitly here (see \cite{depj}) and write 
down the equations of the in-medium gluodynamics using the same method 
as in electrodynamics. We introduce the nuclear
permittivity and denote it also by $\epsilon $, since this will not lead
to any confusion. After that, one should replace ${\bf E}_a$ 
by $\epsilon {\bf E}_a$ and get
\begin{equation}
\label{8}
\epsilon ({\rm div } {\bf E}_a-gf_{abc}{\bf A}_b {\bf E}_c)=\rho _a, \;\;\;\;
{\rm curl } {\bf B}_a-\epsilon \frac {\partial {\bf E}_a}{\partial t} -
gf_{abc}(\epsilon \Phi _b{\bf E}_c + [{\bf A}_b{\bf B}_c])={\bf j}_a.
\end{equation}
The space-time dispersion of $\epsilon $ is neglected here.
 
In terms of potentials these equations are cast in the form
\begin{eqnarray}
\bigtriangleup {\bf A}_a-\epsilon \frac{\partial ^2{\bf A}_a}{\partial t^2}=
-{\bf j}_a -
gf_{abc}(\frac {1}{2} {\rm curl } [{\bf A}_b, {\bf A}_c]+\frac {\partial }
{\partial t}({\bf A}_b\Phi _c)+[{\bf A}_b {\rm curl } {\bf A}_c]-  \nonumber \\
\epsilon \Phi _b\frac 
{\partial {\bf A}_c}{\partial t}- 
\epsilon \Phi _b {\rm grad } \Phi _c-\frac {1}{2} gf_{cmn}
[{\bf A}_b[{\bf A}_m{\bf A}_n]]+g\epsilon f_{cmn}\Phi _b{\bf A}_m\Phi _n), \hfill \label{f.6}
\end{eqnarray}

\begin{eqnarray}
\bigtriangleup \Phi _a-\epsilon \frac {\partial ^2 \Phi _a}
{\partial t^2}=-\frac {\rho _a}{\epsilon }+ 
gf_{abc}(2{\bf A}_b {\rm grad }\Phi _c+{\bf A}_b
\frac {\partial {\bf A}_c}{\partial t}+\epsilon 
\frac {\partial \Phi _b}{\partial t}
\Phi _c)+  \nonumber  \\
g^2 f_{amn} f_{nlb} {\bf A}_m {\bf A}_l \Phi _b. \hfill  \label{f.7}
\end{eqnarray}
If the terms with coupling constant $g$ are omitted, one gets
the set of Abelian equations, that differ from electrodynamical equations
by the color index $a$ only. The external current is due to a parton  
moving fast relative to partons "at rest".

The crucial distinction between the in-vacuum and in-medium equations is that 
there is no radiation (the field strength is zero in the forward light-cone and 
no gluons are produced) in the lowest order solution in vacuum, and it 
is admitted in medium,
because $\epsilon $ takes into account the collective response (color
polarization) of the nuclear matter. 

Cherenkov effects are especially suited for treating them by classical
approach to (\ref{f.6}), (\ref{f.7}). Their unique feature is
independence of the coherence of subsequent emissions on the time interval
between these processes. The lack of balance of the phase $\Delta \phi $ between
emissions with frequency $\omega =k/\sqrt {\epsilon }$ separated by the
time interval $\Delta t $ (or the length $\Delta z=v\Delta t$) is given by
\begin{equation}
\label{f.9}
\Delta \phi =\omega \Delta t-k\Delta z\cos \theta =
k\Delta z(\frac {1}{v\sqrt {\epsilon }}-\cos \theta )
\end{equation}
up to terms that vanish for large distances. For Cherenkov effects the 
angle $\theta $ is
\begin{equation}
\label{f.10}
\cos \theta = \frac {1}{v\sqrt {\epsilon }}.
\end{equation}
The coherence condition $\Delta \phi =0$ is strictly valid independent of 
$\Delta z $. This is a crucial property specific for Cherenkov radiation only.
The fields $(\Phi _a, {\bf A}_a)$ and the classical current for in-medium 
gluodynamics can be represented by the product of the 
electrodynamical expressions $(\Phi , {\bf A})$ and the color matrix $T_a$. 

Let us recall the Abelian solution for the current 
with velocity ${\bf v}$ along $z$-axis:
\begin{equation}
\label{f.11}
{\bf j}({\bf r},t)={\bf v}\rho ({\bf r},t)=4\pi g{\bf v}\delta({\bf r}-{\bf v}t).
\end{equation}

In the lowest order the solutions for the scalar and vector potentials are
related 
${\bf A}^{(1)}({\bf r},t)=\epsilon {\bf v} \Phi ^{(1)}({\bf r},t)$ and
\begin{equation}
\label{f.12}
\Phi ^{(1)}({\bf r},t)=\frac {2g}{\epsilon }\frac {\theta
(vt-z-r_{\perp }\sqrt {\epsilon v^2-1})}{\sqrt {(vt-z)^2-r_{\perp} ^2
(\epsilon v^2-1)}}.
\end{equation}

Here $r_{\perp }=\sqrt {x^2+y^2}$ is the cylindrical coordinate; $z$
symmetry axis. The cone
\begin{equation}
\label{f.14}
z=vt-r_{\perp }\sqrt {\epsilon v^2-1}
\end{equation}
determines the position of the shock wave due to the $\theta $-function
in (\ref{f.12}). The field is localized within this cone and decreases with time 
as $1/t$ at any fixed point. The gluons emission is perpendicular to the cone
(\ref{f.14}) at the Cherenkov angle (\ref{f.10}). 

Due to the antisymmetry of $f_{abc}$, the higher order terms ($g^3$,...)
are equal to zero for any solution multiplicative in space-time and color
as seen from (\ref{f.6}), (\ref{f.7}).

The expression for the intensity of the radiation is given by the Tamm-Frank
formula (up to Casimir operators) that leads to infinity for constant 
$\epsilon $. The $\omega $-dependence of $\epsilon $ (dispersion), its 
imaginary part (absorption) $\epsilon _2$ and chromomagnetic permeability 
can be taken into account \cite{depj}. 

Recently, the experimental data of STAR and PHENIX \cite{ajit, ul} were
fitted \cite{dklv} with account of the imaginary part of $\epsilon $ and 
emission of pions and $\rho $-mesons within the Cherenkov cone.
The results are presented in Table 1 (for more details see \cite{dklv}).

\begin{center}
{\bf Table 1}

\bigskip

\begin{tabular}{|c|c|c|c|}
  \hline
  Experiment & $\theta_{\rm max}$ & $\varepsilon_1$ & $\varepsilon_2$   \\ \hline
  STAR & 1.04~rad & 3.95 & 0.8   \\ \hline
  PHENIX & 1.27~rad & 9.5 & 1.8   \\ \hline
\end{tabular}
\end{center}

The real parts $\epsilon _1$ are quite large while the imaginary parts are
small so that $(\epsilon_2/\epsilon _1)^2\approx 0.04 \ll 1$. Different 
values of $\epsilon _1$ for STAR and PHENIX are related to
different positions of hump maxima in these experiments.

The theoretical attempts to estimate the nuclear permittivity from first 
principles are not very convincing \cite{kk, we, bi, ko, dpri}. 
Therefore, I prefer to use the general formulae of the scattering theory for
the nuclear permittivity. It is related to the refractive index $n$ 
of the medium $\epsilon =n^2$
and the latter one is expressed \cite{gw} through the real part of the forward 
scattering amplitude of the refracted quanta ${\rm Re}F(0^{\rm o},E)$ by
\begin{equation}
\label{f.19}
{\rm Re} n(E )=1+\Delta n_R =1+\frac {6m_{\pi }^3\nu }{E^2}{\rm Re }F(E) =
1+\frac {3m_{\pi }^3\nu }{4\pi E } \sigma (E )\rho (E ).  
\end{equation}
Here $E$ denotes the energy, $\nu $ the number of scatterers within a single 
nucleon, $m_{\pi }$ the pion mass, $\sigma (E)$ the cross section and $\rho (E)$ 
the ratio of real to imaginary parts of the forward scattering amplitude $F(E)$. 

Thus the emission of Cherenkov gluons is possible only for processes with 
positive ${\rm Re} F(E)$ or $\rho (E)$. Unfortunately, we are unable to 
calculate directly in QCD these characteristics of gluons
and have to rely on analogies and on our knowledge of the properties of hadrons. 
The only experimental facts we get for this medium are brought about 
by particles registered at the final stage. They have some features in common, 
which (one may hope!) are also relevant for gluons as the carriers of the 
strong forces. Those, first, are the resonant behavior of amplitudes at rather 
low energies and, second, the positive real part of the
forward scattering amplitudes at very high energies for hadron-hadron and
photon-hadron processes as measured from the interference of the Coulomb and
hadronic parts of the amplitudes. ${\rm Re} F(0^{\rm o},E)$ is always positive 
(i.e., $n>1$) within the low-mass wings of the Breit-Wigner resonances.  
This shows that the necessary condition for Cherenkov effects $n>1$ is
satisfied at least within these two energy intervals. This fact was used
to describe experimental observations at SPS, RHIC and cosmic ray energies. 
The asymmetry of
the $\rho $-meson shape at SPS \cite{da} and azimuthal correlations of 
in-medium jets at RHIC \cite{ul, ajit} were explained by emission of 
comparatively low-energy Cherenkov gluons \cite{dnec, drem1}.
The parton density and intensity of the radiation were estimated. In its turn,
cosmic ray data \cite{apan} at energies corresponding to LHC require very 
high-energy gluons to be emitted by the ultrarelativistic partons moving along 
the collision axis \cite{d1}. The specific predictions at LHC stemming from this
observation were discussed elsewhere \cite{drem2}. Let us note the important 
difference from electrodynamics, where $n<1$ at high frequencies. The energy
of the forward moving partons at LHC would exceed the thresholds above which 
$n>1$. Then both types of experiments can be done, i.e. the 90$^{\rm o}$-trigger
and non-trigger forward-backward partons experiments. The predicted results for 
90$^{\rm o}$-trigger geometry are similar to those at RHIC. The non-trigger
Cherenkov gluons should be emitted within the rings at polar angles of tens
degrees in c.m.s. at LHC by the forward moving partons (and symmetrically by 
the backward ones) according to some events observed in cosmic
rays \cite{apan, drem1}. This is the new prediction for LHC.

To conclude, the in-medium gluodynamics leads quite naturally to the prediction 
of Cherenkov gluons emitted within the nuclear medium if $\epsilon >1$. The 
experimental data about the nimbus of away-side jets obtained at RHIC have 
been well fitted by these formulae with complex nuclear permittivity. Quite 
large values of its real part are estimated from fits to experimental data. 
Therefrom one concludes that the density of scatterers $\nu $ is rather high
(about 10-20 per a hadron). The imaginary part is 
comparatively small. The specific predictions at LHC energies are waiting
for their verification. 
 
{\it Acknowledgement.} This work is supported by RFBR grants 06-02-17051, 
06-02-16864, 08-02-91000-CERN.

\end{document}